\documentclass[reprint,showpacs,showkeys,amsmath,amssymb,twocolumn,aps,superscriptaddress,floatfix]{revtex4-2}
\usepackage{amssymb} 
\usepackage{amsmath,bm} 
\usepackage{graphicx} 
\usepackage[normalem]{ulem} 
\usepackage{multirow} 
\usepackage[colorlinks,linkcolor=blue,urlcolor=blue,citecolor=blue]{hyperref} 
\usepackage{lipsum} 
\usepackage[usenames, dvipsnames]{xcolor} 
\usepackage{tensor} 
\usepackage{isotope} 
\usepackage{amsmath} 
\usepackage{url} 
\allowdisplaybreaks[4] 

\renewcommand{\sout}{\bgroup \color{red} \ULdepth=-.5ex \ULset}

\begin{document}  

\title{Investigating $^{238}$U Deformation via Dilepton Production in Relativistic Heavy-Ion Collisions} 

\author{Wen-Hao Zhou}
\email{zhouwenhao@xaau.edu.cn}
\affiliation{Key Laboratory of Nuclear Physics and Ion-beam Application (MOE), Institute of Modern Physics, Fudan University, Shanghai, 200433, China}
\affiliation{Faculty of Science, Xihang University, Xi‘an, 710077, China}
\affiliation{Shanghai Research Center for Theoretical Nuclear Physics, NSFC and Fudan University, Shanghai 200438, China}

\author{Lu-Meng Liu}
\email{liulumeng@fudan.edu.cn}
\affiliation{Shanghai Research Center for Theoretical Nuclear Physics, NSFC and Fudan University, Shanghai, 200438, China}

\author{Che Ming Ko}
\email{ko@comp.tamu.edu}
\affiliation{Cyclotron Institute and Department of Physics and Astronomy, Texas A\&M University, College Station, Texas 77843, USA}

\author{Kai-Jia Sun}
\email[Corresponding author: ]{kjsun$@$fudan.edu.cn}
\affiliation{Key Laboratory of Nuclear Physics and Ion-beam Application (MOE), Institute of Modern Physics, Fudan University, Shanghai, 200433, China}
\affiliation{Shanghai Research Center for Theoretical Nuclear Physics, NSFC and Fudan University, Shanghai 200438, China}

\author{Jun Xu}
\email{junxu@tongji.edu.cn}
\affiliation{School of Physics Science and Engineering, Tongji University, Shanghai, 200092, China}

\date{\today}

\begin{abstract}
	Due to their weak coupling to the strongly interacting matter produced in relativistic heavy-ion collisions, dileptons serve as a sensitive probe of the initial geometry of the colliding nuclei. In this study, we investigate the influence of initial nuclear quadrupole deformation, characterized by the parameter $\beta_2$, on dilepton production in $\text{U}+\text{U}$ collisions at $\sqrt{s_{NN}}=193$ GeV. The analysis is carried out using a modified multiphase transport model in which partonic interactions are described by the Nambu–Jona-Lasinio model. We observe a clear linear dependence of dilepton yields on $\beta_2^2$ in both the low-mass region (LMR, $<1\,\mathrm{GeV}\!/c^2$) and intermediate-mass region (IMR, $1-3\,\mathrm{GeV}\!/c^2$) of the dilepton spectrum for the most central collisions. Also, dilepton production in the IMR region exhibits a stronger sensitivity to nuclear deformation than in the LMR, reflecting the dominance of earlier partonic processes in this mass range. These results suggest that precise measurements of dilepton yields in relativistic heavy-ion collisions can provide a viable means to determine the deformation parameter $\beta_2$ of $^{238}$U.
		\end{abstract} 
\maketitle	
	\section{Introduction}
	Nuclear deformations can be understood as a quantum phenomenon arising from spontaneous symmetry breaking in finite systems~\cite{osti_4091234,ring2004nuclear,Guidry:2022vyd}. Investigating nuclear deformation contributes to a deeper understanding of fission dynamics~\cite{Goddard:2015eta}, $r$-process nucleosynthesis~\cite{Goriely:1998utv}, and nuclear reactions in evolving stars~\cite{Thielemann:2023oou}. In particular, understanding the deformation of $^{238}$U is important for clarifying geometric-driven background effects in experimental measurements of the chiral magnetic effect (CME)~\cite{Fukushima:2008xe,Voloshin:2010ut,Deng:2016knn,Sun:2018idn,Zhao:2022grq,Yuan:2023skl,Yuan:2024wpz}.
	
	Nuclear deformation has traditionally been determined by measuring the reduced electric transition probabilities in a variety of low-energy nuclear experiments~\cite{Cline:1986ik,Yang:2022wbl}. Recently, imaging the shapes, including deformation~\cite{Jia:2021tzt,Jia:2021oyt,Jia:2021wbq,Zhang:2021kxj,Ryssens:2023fkv,Wang:2024vjf,Giacalone:2019pca,Giacalone:2021uhj,Bally:2021qys,Zhao:2024lpc,STAR:2024wgy}, neutron skin thickness~\cite{Liu:2023qeq,Ding:2024xxu,Liu:2022kvz,Giacalone:2023cet,Liu:2023pav}, and cluster structures~\cite{Zhang:2024vkh,Zhao:2024feh,Wang:2024ulq,Liu:2023gun,Giacalone:2024luz,YuanyuanWang:2024sgp,Liu:2025zsi}, via relativistic heavy-ion collisions (HICs) has attracted considerable interest. In particular, recent experiments on high-energy $\mathrm{U}+\mathrm{U}$ collisions at $\sqrt{s_{NN}}=193$ GeV offer a valuable opportunity to extract the quadrupole deformation parameter $\beta_2$ of the colliding nuclei. The reported experimental value, $0.286\pm 0.025$~\cite{STAR:2024wgy}, is consistent with earlier measurements from lower-energy experiments~\cite{Pritychenko:2013gwa} such as $0.287\pm0.007$. 
	
	Beyond this direct measurement, previous studies of deformed nuclei in HICs primarily utilized observables such as collective flow, transverse momentum fluctuations, and their correlations~\cite{Jia:2021tzt,Jia:2021oyt,Zhang:2021kxj,Jia:2021wbq}. These approaches are based on the premise that spatial anisotropies in the initial nuclear geometry, arising from deformation, are converted into final-state momentum anisotropies due to strong pressure gradients during the partonic and hadronic phases~\cite{Jia:2021tzt,Jia:2021oyt,Zhang:2021kxj,Jia:2021wbq}. The value of $\beta_2$ is typically extracted by exploiting polynomial relationships between these observables and $\beta_2$, and often by comparing scaling behaviors with those in reference systems with known deformation parameters~\cite{Jia:2021tzt,STAR:2024wgy}.
	
	Unlike flow-related probes, which can be influenced by non-flow effects such as resonance decays and hadronic rescattering~\cite{Bierlich:2021poz}, dilepton production provides more direct information about the initial stage of heavy-ion collisions. This is because dileptons interact only weakly with the surrounding hadronic medium. In particular, dileptons in the intermediate-mass region (IMR) of the excess invariant spectrum are predominantly produced through quark-antiquark annihilation in the quark-gluon plasma (QGP), a process that occurs primarily in the early stage of the collision~\cite{STAR:2023wta}. Numerous experimental measurements of dilepton production have been reported~\cite{STAR:2015tnn,STAR:2015zal,Seck:2021mti,STAR:2023wta,PHENIX:2015vek,STAR:2018ldd,Specht:2010xu,ALICE:2018ael,HADES:2019auv,Chen:2024aom}, and dileptons in different invariant mass regions are commonly used to extract the temperature of both the partonic and hadronic phases. Furthermore, Ref.~\cite{Luo:2023syp} investigated the impact of $^{238}\text{U}$ deformation on dilepton photoproduction in ultra-peripheral collisions (UPCs) and found an approximately $3\%$ difference in yields between spherical and deformed nuclear configurations.
	
	In the present study, we investigate the effect of initial nuclear quadrupole deformation $\beta_2$ on dilepton production in $\text{U}+\text{U}$ collisions at $\sqrt{s_{NN}}=193$ GeV, using a modified multiphase transport model in which partonic interactions are governed by the Nambu–Jona-Lasinio (NJL) model~\cite{Ko:2013nbt,Ko:2012lhi,Xu:2013sta}. To reduce the influence of volume effects, (i.e., variations in fireball size), finite rapidity cuts (i.e., detector acceptance limitations), and model dependence, we normalize the dilepton yield $N_{ll}$ by the charged particle multiplicity $N_\mathrm{ch}$. Using dilepton production in Au+Au collisions as a reference, we construct the following double yield ratio:
	\begin{equation}
		R_\mathrm{U-Au} = \frac{(\mathrm{d}N_{ll}^\mathrm{U}/\mathrm{d}y)/
			(\mathrm{d}N_\mathrm{ch}^\mathrm{U}/\mathrm{d}y)}
		{(\mathrm{d}N_{ll}^\mathrm{Au}/\mathrm{d}y)/
			(\mathrm{d}N_\mathrm{ch}^\mathrm{Au}/\mathrm{d}y)}. \label{equ:ratio}
	\end{equation}
    This ratio is primarily sensitive to the difference in nuclear deformation between U and Au. We find a clear linear correlation between $\beta_2^2$ and $R_\text{U-Au}$ in both the low-mass region (LMR, $<1\,\mathrm{GeV}\!/c^2$), where dilepton yields are dominated by $\rho^0$ meson decays in the hadronic phase, and the intermediate-mass region (IMR, $1-3\,\mathrm{GeV}\!/c^2$), where dileptons are predominantly produced via quark-antiquark annihilation in the partonic phase of the collision. Moreover, we find that dilepton production in the IMR exhibits a stronger sensitivity to the deformation parameter $\beta_2$ of $^{238}$U compared to the LMR, as IMR dileptons originate primarily from the early-stage partonic phase. 
	
	\section{Method}
	
	The present study is based on a multiphase transport (AMPT) model, with the partonic transport evolution replaced by the NJL transport model~\cite{Ko:2013nbt,Ko:2012lhi,Xu:2013sta}. The AMPT model incorporates initial conditions from the Heavy Ion Jet Interaction Generator (HIJING)~\cite{Wang:1991hta,Gyulassy:1994ew}, which includes fragmentation and jet production, along with a string melting mechanism that converts exciting strings into partons. Quark coalescence is used to hadronize partons at freeze out, and the subsequent hadronic rescattering is described by a relativistic transport (ART) model~\cite{Li:1995pra}, which includes various resonance decay modes as well as elastic and inelastic hadron-hadron scatterings. In addition, we implement the initial-state density distributions for deformed nuclei within the AMPT framework and include dilepton production channels in both the partonic and hadronic phases. These components will be discussed in more detail below.
	
	\begin{figure}[htbp]
		\centering
		\includegraphics[width=0.48\textwidth]{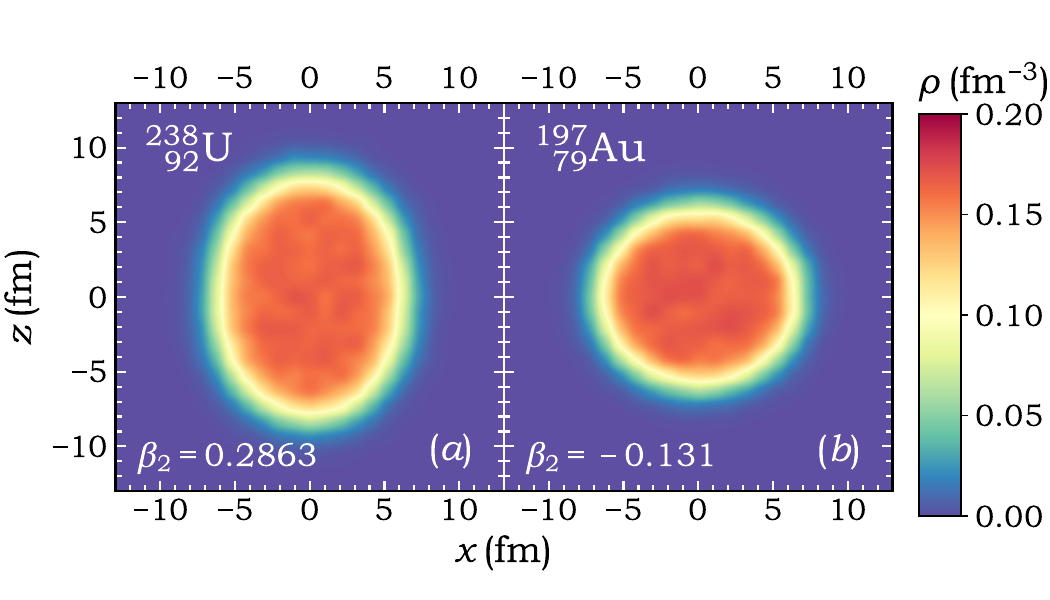}
		\caption{Nucleon density distributions of $^{238}\mathrm{U}$ and $^{197}\mathrm{Au}$ nuclei within $|r_y|<0.5$ fm. The values of $\beta_2$ are taken from Ref.~\cite{Raman:2001nnq}.}
		\label{fig:structureAuU}
	\end{figure}

	In HICs, the nuclear density of the deformed $^{238}$U nucleus is typically modeled using the Woods-Saxon distribution~\cite{Giacalone:2020awm}:
	\begin{equation}
		\rho(r,\theta,\phi) = \frac{\rho_0}{1+\exp{\left[\frac{r-R_\mathrm{WS}(1+\beta_2Y_{2,0}(\theta,\phi))}{a_\mathrm{WS}}\right]}}.
	\end{equation}
	Here, $r$, $\theta$ and $\phi$ represent the radial distance, polar angle and azimuthal angle, respectively. The parameter $a_\mathrm{WS}$ denotes the surface thickness, $R_\mathrm{WS}$ is the nucleus radius, $\beta_2$ characterizes the quadrupole deformation, and $Y_{2,0}(\theta,\phi)$ is the second-order spherical harmonic function. Figure~\ref{fig:structureAuU} shows the nucleon distributions of the deformed $^{197}\mathrm{Au}$ and $^{238}\mathrm{U}$ nuclei. The deformed Woods-Saxon parameters used are taken from Ref.~\cite{Raman:2001nnq}: $R_\mathrm{WS}=6.8054$ fm, $a_\mathrm{WS}=0.605$ fm and $\beta_2=0.2863$ for $^{238}\mathrm{U}$; and $R_\mathrm{WS}=6.38$ fm, $a_\mathrm{WS}=0.535$ fm and $\beta_2=-0.131$ for $^{197}\mathrm{Au}$. As shown, both nuclei exhibit axial asymmetry with respect to the $z$-direction, with $^{238}\mathrm{U}$ displaying a prolate shape and $^{197}\mathrm{Au}$ an oblate shape.
	
	The evolution  of the partonic phase is described by the NJL model, and the Lagrangian density of three-flavor NJL model used in this study is given by~\cite{Nambu:1961tp,Nambu:1961fr,Buballa:2003qv}
	\begin{align}
		\mathcal{L}_{\mathrm{NJL}} 
		&= \bar{\psi}(i\gamma^\mu\partial_\mu-\hat{m})\psi\notag\\
		&+\frac{G_{S}}{2}\sum_{a=0}^{8}[(\bar{\psi}\lambda_a\psi)^2+
		(\bar{\psi}i\gamma_5\lambda_a\psi)^2]
		\notag\\
		&-K\{\mathrm{det}_f[\bar{\psi}(1+\gamma_5)\psi]+
		\mathrm{det}_f[\bar{\psi}(1-\gamma_5)\psi]\},
	\end{align}
	where $\psi=\mathrm{diag}\left(u,d,s\right)$ represents the 3-flavor quark fields, and $\hat{m} = \mathrm{diag}\left(m_u,m_d,m_s\right)$ denotes the current quark mass matrix in flavor space. The matrices $\lambda_{1-8}$ are the standard Gell-Mann matrices, and $\lambda_0=\sqrt{2/3}I$ is proportional to the identity matrix.  The constants $G_S$ and $K$ represent the strengths of the scalar coupling and the Kobayashi-Maskawa-t’Hooft (KMT) interaction, respectively. In this study, we adopt the parameter values $m_u=m_d=5.5$ MeV, $m_s=140.7$ MeV, $G_S\Lambda^2=3.67$, $K\Lambda^5=12.36$, and a momentum cutoff $\Lambda=602.3\,\mathrm{MeV\!}/c$ for the integrals, as given in Refs.~\cite{Buballa:2003qv,Hatsuda:1994pi}. 
	
	The model accommodates a first-order phase transition and has been used to investigate the spinodal instabilities in the partonic phase~\cite{Sun:2020bbn,Sun:2020pjz,Li:2016uvu,Sun:2022cxp}. It has also been applied to describe collective flow phenomena~\cite{Xu:2013sta,Guo:2018kdz,Liu:2019ags,Zhou:2021ruf} through the inclusion of vector and/or isovector couplings. However, these effects are weak and thus neglected in this study, as we focus on the high-temperature, quark-antiquark symmetric matter produced at the top RHIC energy.
	
	Under the mean-field and semi-classical approximations, the single-particle Hamiltonian for a quark of flavor $i$ in the NJL model is given by $H_i = \sqrt{M_i^2+\vec{k}_i^2}$~\cite{Ko:2013nbt}, where $\vec{k}_i$ is the kinetic momentum. The corresponding equations of motion (EOMs) for quarks, derived from the canonical equations, are:
	\begin{equation}
		\frac{\mathrm{d}\vec{r}_i}{\mathrm{d}t}=\frac{\vec{k}_i}{E_i},\quad
		\frac{\mathrm{d}\vec{k}_i}{\mathrm{d}t}=-\frac{M_i}{E_i}\vec{\nabla}M_i,
	\end{equation}
	where $M_i$ and $E_i=\sqrt{M_i^2+\vec{k}_i^2}$ represent the effective mass and energy of partons, respectively. A stochastic collision criterion is employed to simulate elastic parton scatterings, and the test-particle method is used to obtain the quark and antiquark phase-space distributions in the transport model simulation~\cite{Wong:1982zzb}. Isotropic elastic scattering with a fixed cross section of $3$ mb is adopted throughout this work. To ensure that partons always follow the Fermi-Dirac distributions, Pauli blocking is implemented.
	
	In experimental analyses, the dilepton contributions from $\rho^0$ meson and the QGP are typically extracted from background sources using so-called cocktail simulations. In our model, we include only these two dilepton production channels and refer them as the excess dileptons~\cite{Linnyk:2011vx}. 
	
	In the partonic phase, dileptons are primarily produced via quark-antiquark annihilation, $q\bar q\rightarrow e^+e^-$. The leading-order cross section for this process is given by~\cite{Halzen:1984mc}
    \begin{equation}	
        \sigma=\frac{4\pi\alpha^2e_i^2}{9s}, 
    \end{equation}
    where $s$ is the invariant mass squared, $e_i$ is the quark electric charge, and $\alpha$ is the fine-structure constant.

    In the hadronic phase, we include in-medium modifications of the $\rho^0$ meson,  which affect both its mass and width (see Ref.~\cite{Li:1994cj} for details). The dominant channel for $\rho^0$ production is $\pi^+ + \pi^- \rightarrow \rho^0$, with the cross section taken from Refs.~\cite{Li:1994cj,Cassing:1999es}. Additionally, $\rho^0$ mesons can be formed through quark coalescence during hadronization. Due to the relatively low production rate of dileptons from both $\rho^0$ decays and quark-antiquark annihilation, these channels are treated perturbatively in our simulation~\cite{Li:1994cj,Linnyk:2015rco}.
	
	\section{Results and discussions}
	
	\begin{figure}[htbp]
		\centering
		\includegraphics[width=0.45\textwidth]{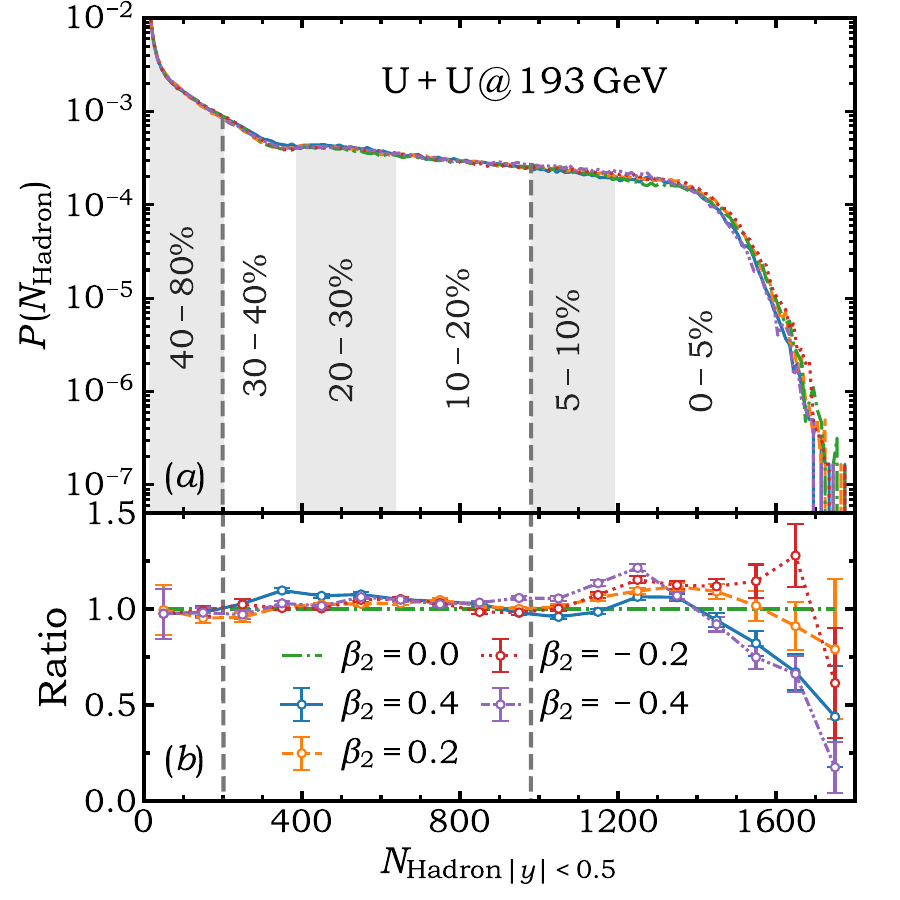}
		\caption{Multiplicity distributions of $^{238}\mathrm{U}$ for different deformation parameters $\beta_2$.}
		\label{fig:multiURandom}
	\end{figure}	
	
	Figure~\ref{fig:multiURandom} shows the normalized hadron multiplicity distributions for five different values of the deformation parameter $\beta_2$. The borders of the centrality bins are defined by using the $\beta_2=0$ case. In the most central collisions ($0-5\%$), larger $|\beta_2|$ leads to greater suppression in multiplicity compared to the spherical nucleus case ($\beta_2=0$). This is because more deformed nuclei typically have a smaller overlap volume, reducing the number of participant nucleons in the collision.
	
	\begin{figure}[htbp]
		\centering
		\includegraphics[width=0.40\textwidth]{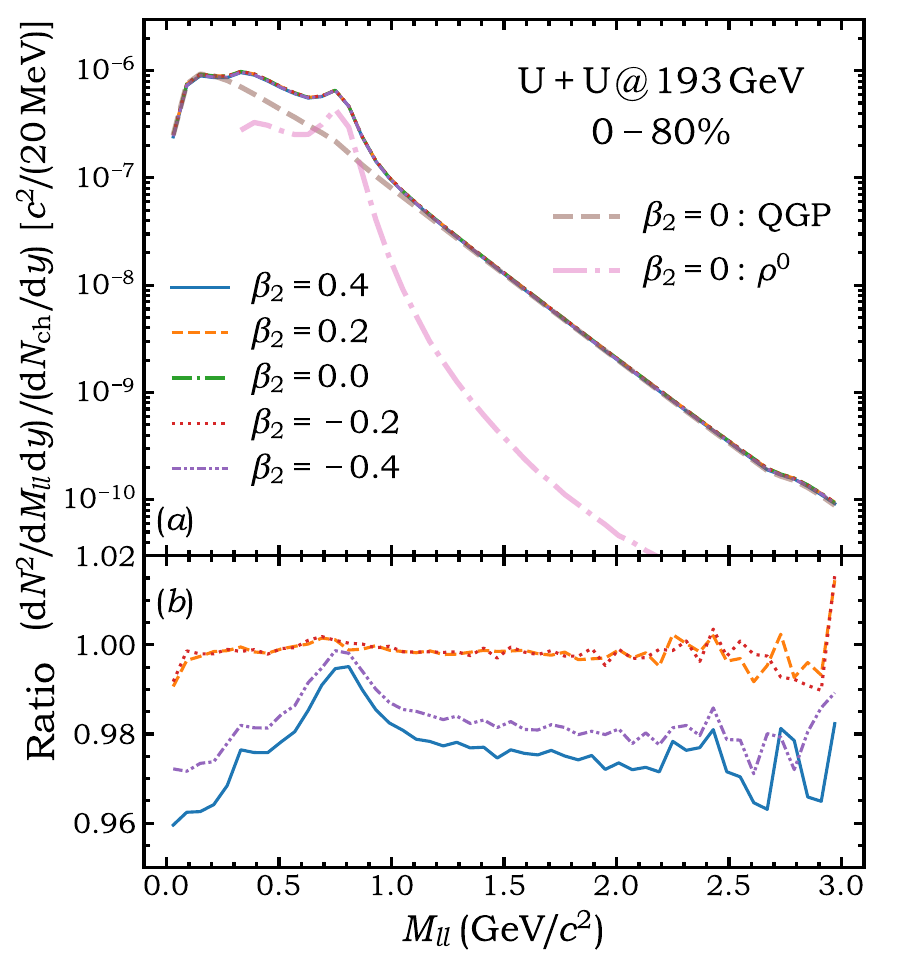}
		\caption{Invariant mass spectra of dileptons in U+U collisions at $\sqrt{s_{NN}}=193$ GeV for different deformation parameters $\beta_2$ $(a)$, and the corresponding ratios relative to the $\beta_2=0$ case $(b)$. The contribution from QGP and $\rho^0$ mesons for the $\beta_2=0$ case are also shown.}
		\label{fig:dileptonUURandom_dNdy}
	\end{figure}
	
	Panel $(a)$ of Fig.~\ref{fig:dileptonUURandom_dNdy} presents the dilepton invariant mass spectra, normalized by the number of charged particles, for different $\beta_2$ values. The centrality selection, based on the hadron multiplicity within $|y|<0.5$, follows that shown in Fig.~\ref{fig:multiURandom}. While the overall shapes of the normalized spectra are similar across different $\beta_2$ values, their ratios to the $\beta_2=0$ case, shown in panel $(b)$, reveal a clear trend: larger $|\beta_2|$ results in stronger suppression of the dilepton yield. This is attributed to the smaller overlap volume due to the random spatial orientation of the nuclear symmetry axis. A smaller overlap region implies fewer produced partons, and thus a lower dilepton yield. 
	
	The dilepton yields from QGP or $\rho^0$ mesons in the spherical case are also shown. Around the $\rho^0$ vacuum pole mass of 775.26 $\mathrm{MeV}\!/\!c^2$, the $\rho^0$ contribution dominates, and the yield ratios are close to unity. This indicates that  $\rho^0$ meson decay is relatively insensitive to $\beta_2$. Since the freeze-out condition is kept the same for different $\beta_2$ values, the final parton distributions are similar, even though the partonic lifetimes differ slightly. As a result, the production of $\rho$- and $\pi$-mesons at hadronization, and thus the hadronic-phase dilepton yield, is largely unaffected by the initial nuclear deformation. Outside this region, the QGP contribution becomes dominate and shows a clear dependence on $\beta_2$, i.e., larger $|\beta_2|$ values result in fewer dileptons.
	
	\begin{figure}[htbp]
		\centering
		\includegraphics[width=0.45\textwidth]{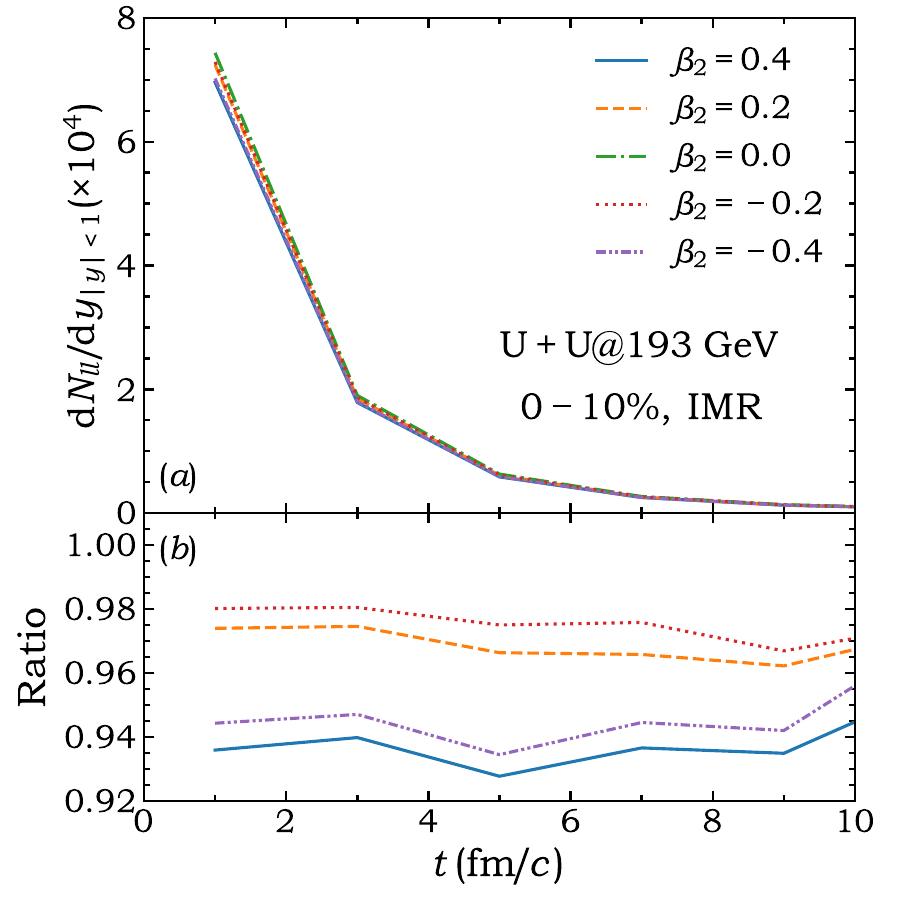}
		\caption{Time evolution of the dilepton yield in the IMR, showing only the QGP contribution, for different deformation parameters $\beta_2$.}
		\label{fig:fireball_time2}
	\end{figure}
	
	Figure~\ref{fig:fireball_time2} shows the time evolution of the dilepton yield in IMR, along with the yield ratio between deformed and spherical nuclei. Most dileptons are  produced in the early stage of the collisions, and the yield decreases rapidly with time. Panel $(b)$ of Fig.~\ref{fig:fireball_time2} shows that larger $|\beta_2|$ consistently results in stronger suppression, independent of the evolution time.
	
	\begin{figure}[htbp]
		\centering
		\includegraphics[width=0.45\textwidth]{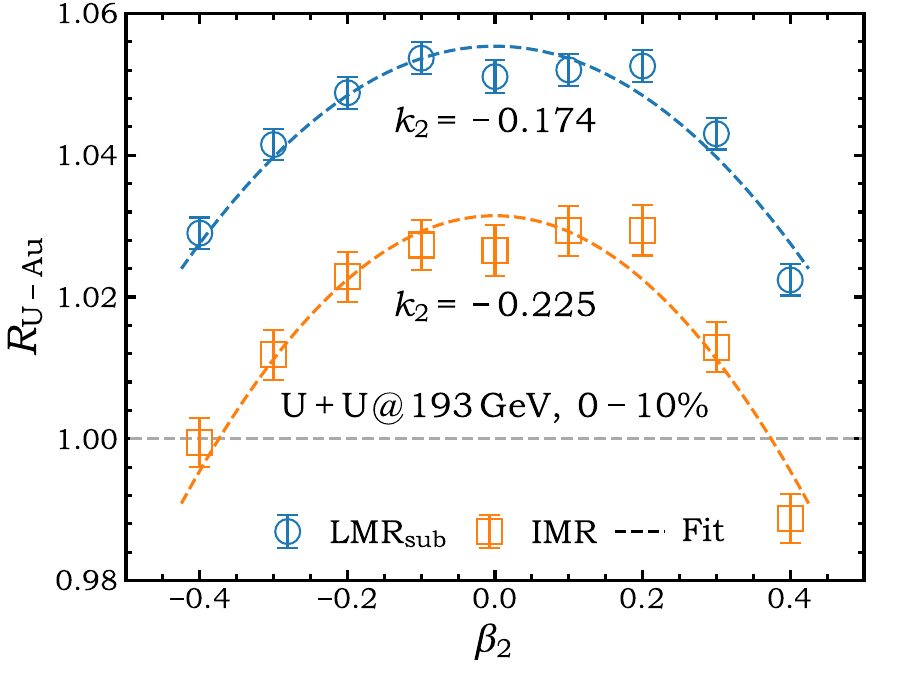}
		\caption{Ratio of the integrated invariant mass spectra of dileptons in U+U collisions at $\sqrt{s_{NN}}=193$ GeV to those in Au+Au collisions at $\sqrt{s_{NN}}=200$ GeV collision as a function of $\beta_2$ in the most central collisions. The dashed lines represent the fitted curves.}
		\label{fig:ratio_beta}
	\end{figure}
	
	Figure~\ref{fig:ratio_beta} exams the effect of nuclear deformation on the ratio $R_\mathrm{U-Au}$ (as defined in Eq.~\eqref{equ:ratio}) in two invariant mass windows: $0.4<M_{ll}<0.75\,\mathrm{GeV}$ (a subregion of the LMR, denoted as LMR$_\mathrm{sub}$) and $1.0<M_{ll}<3.0\,\mathrm{GeV}$ (denoted as IMR) in the most central collisions. The deformation parameter for the Au nuclei is taken as $\beta_2=-0.131$~\cite{Raman:2001nnq}, though using $\beta_2=0$ for Au yields less than $1\%$ difference in the dilepton spectrum. The results show a linear dependence of $R_\mathrm{U-Au}$ on $\beta^2_2$, which can be written as
	\begin{equation}
		R_\mathrm{U-Au} = k_0 + k_2 \beta^2_2, \label{equ:fit}
	\end{equation}
	where $k_0$ is the intercept and $|k_2|$ indicates the sensitivity to deformation. As previously shown~\cite{Jia:2021oyt}, larger $\beta_2$ leads to suppression in hadron multiplicity, corresponding to fewer participant nucleons. 
	
	Figure~\ref{fig:ratio_beta} also presents the fitted values of $k_2$ for both spectral regions. The value of $|k_2|=0.174$ in LMR$_\mathrm{sub}$ is smaller than $|k_2|=0.225$ in the IMR.  This is because LMR$_\mathrm{sub}$ dileptons are mostly from $\rho^0$ decay,  which is less sensitive to the number of participant nucleons, while IMR dileptons originate mainly from the QGP and thus reflect parton dynamics more directly.
	
	The quadratic dependence in Eq.~\eqref{equ:fit} can be understood as follows. In relativistic HICs (i.e., $\sqrt{s_{NN}}\geqslant100$ GeV), the numbers of particles and antiparticles are nearly equal~\cite{STAR:2008med}, implying $N_q\approx N_{\bar q}$ in central events. The dilepton yield from QGP is approximately proportional to the product $N_{ll} \propto N_q N_{\bar q}$. After hadronization, the partons convert to hadrons that are predominantly mesons at this energy scale, so $N_\mathrm{hadron}\propto N_\mathrm{parton}$. Prior work~\cite{Jia:2021wbq} has shown that the hadron multiplicity scales linearly with $\beta_2^2$~\cite{Jia:2021wbq}, i.e., $N_\mathrm{hadron} = a_0 + b_0\beta^2_2$. Assuming $ N_\mathrm{ch}\propto N_\mathrm{hadron}$, the dilepton yield becomes 
    \begin{equation}
        N_{ll}\propto N_\text{parton}^2\propto N_\text{hadron}^2\propto N_\text{ch}^2\propto(a_0+b_0\beta_2^2)^2.
    \end{equation}
   We therefore have $N_{ll}/N_\text{ch}\propto a_0+b_0\beta_2^2$, which explains the quadratic dependence in Eq.~\eqref{equ:fit}.
	
	\section{Summary}
	
	In this study, we investigate the effects of nuclear deformation on dilepton production using the NJL-extended AMPT model. We find the dilepton yield scales with $\beta_2^2$, with the ratio $R_\mathrm{U-Au}$ exhibiting a clear dependence on $\beta_2^2$ in both LMR$_\mathrm{sub}$ and IMR, but largely insensitive to the sign of $\beta_2$. Moreover, the slope $|k_2|=0.225$ extracted in the IMR is larger than the slope $|k_2|=0.174$ in the LMR$_{\rm sub}$, indicating that dilepton from the QGP is more sensitive to deformation than that from $\rho^0$ meson decays. These findings suggest that the dilepton yield ratio can serve as a sensitive probe of the deformation parameter of $^{238}$U and provide valuable guidance for future experimental studies.
	
	Our approach can also be extended to relativistic isobar collisions to extract the $\beta_2$ values of isobar nuclei such as $^{96}$Ru and $^{96}$Zr. In addition to probing nuclear deformation, the temperature of the QGP can be extracted from the dilepton invariant mass spectra~\cite{Rapp:2014hha,Zhou:2024yyo}. In future work, we aim to extract the temperature and other thermal properties of QGP from high-energy heavy-ion collisions, through a combination of experimental data~\cite{STAR:2015tnn,STAR:2015zal,Seck:2021mti,STAR:2023wta,PHENIX:2015vek,STAR:2018ldd,Specht:2010xu,ALICE:2018ael,HADES:2019auv,Chen:2024aom} and further model calculations.
	
    \section*{Acknowledgments}
	The authors thank Wang-Mei Zha, Zao-Chen Ye and Shuai Yang for helpful discussions. This work is supported in part by the National Natural Science Foundation of China (NSFC) Grant Nos: 12422509, 12147101, 12347106, 12347143, 12375121, 12375125, and 12405150; the China Postdoctoral Science Foundation under Grant No. 2024M750489; the Fundamental Research Funds for the Central Universities; and the U.S. Department of Energy under Award No. DE-SC$0015266$. The computations in this research were performed using the CFFF platform at Fudan University.
	
	\bibliographystyle{apsrev4-2}

%
	
\end{document}